\documentclass[a4paper,10pt]{article}
\usepackage[utf8]{inputenc}
\usepackage{authblk} %For proper title page with authors
\usepackage{amsmath, amsthm, amsfonts, amssymb} % for advanced maths
\usepackage{mathrsfs} % for calligraphic 
\usepackage{bbm}
\usepackage{mathtools}
\usepackage{doi}
\usepackage{hyperref} %For hyperlinks
\usepackage{braket}
\usepackage[backend=bibtex,sorting=none]{biblatex}
\renewbibmacro{in:}{}
\DeclareFieldFormat{pages}{#1}
\usepackage{IKKtex} %all newcommands are located in separate package.

\numberwithin{equation}{section}% numbering equations to each section

\allowdisplaybreaks

\title{\Huge The thermal backreaction of a scalar field in de Sitter spacetime. II. \\  Spectrum enhancement and holography}
\author[1]{Antonis Kalogirou\thanks{akalogirou@mail.ntua.gr}}
\affil[1]{\textit{Department of Physics, School of Applied Mathematical and Physical Sciences, National Technical University of Athens, Zografou Campus, GR-15780 Athens, Greece} \newline}

\date{}

\begin{document}
\maketitle

\vskip 1.5 true cm
\centerline {\bf Abstract}
\vskip 3.0ex
%\thintablerule
\vskip 2.0ex
%\vskip 1.0ex\noindent

We study a spacetime obtained from the semi-classical backreaction computed via the Thermo-field dynamics approach in the Poincare patch of de Sitter spacetime. The resulting bulk equation takes the Whittaker form and we examine two distinct applications. At leading order, the co-moving curvature perturbations are shown to match a constant-roll model in the frozen attractor regime, corresponding to a UV enhancement of the spectrum with $n_S \sim 2$. This blue tilt arises only for modes exiting the horizon during a transient late-time phase of inflation and therefore does not affect perturbations in the CMB scale. In the holographic context, we compute the CFT two-point function at the future boundary, and away from it we construct the flow-equation of the dual QFT that matches the beta-function of the $Sp(N)$ model in three dimensions.

\vskip 2.0ex
%\thintablerule

\vskip-0.2cm
\newpage

\newpage

%\tableofcontents
%
%\newpage

\section{Introduction}
In early-universe cosmology, it is suggested that the universe underwent a brief epoch of rapid expansion, commonly referred to as cosmic inflation. The latter is successfully described by a spacetime metric in Friedmann-Lemaître-Robertson-Walker (FLRW) coordinates, characterized by
a scale factor $a(t)$ which relates physical distances to the comoving spatial coordinates $x^i, \ i=1,...,d$. In the presence of a cosmological constant $\Lambda$ driving the expansion, the assumptions of homogeneity and isotropy lead to the solution in cosmic time $t$
\bad
a_0 (t) = e^{ H_0 t }
\ead 
where $\Lambda = \frac{d(d-1)}{2} H^2_0$. In these coordinates, the metric takes the form of the Poincaré (co-moving) patch of de Sitter (dS) spacetime which we will restrict to 4 ($d=3$) dimensions. Furthermore, dS spacetime is maximally symmetric, with its isometry group $SO(1,4)$ constraining the structure of quantum correlation functions. This makes it a particularly useful toy model for studying quantum fields and semiclassical gravitational effects in curved spacetime. Therefore, one usually considers quantum fields at a fixed curved background which for negligible gravitational quantum effects, is considered as a good first order approximation. 

Spacetimes with non-zero curvature in most cases lack the notion of a global timelike Killing vector. Consequently, there is no unique way of defining positive-frequency modes since they are observer-dependent and any initial choice can be written as a linear combination of mixed modes via a Bogolyubov transformation (BT). Similarly to the case of black holes, \cite{GibbonsHawking} showcased that a timelike observer, carrying an Unruh detector in a spacetime containing an event horizon, should measure a thermal density of particles characterized by $\beta_{\rm dS} \equiv 1 / T_{\rm dS} = 2 \pi / H_0 $ which we define as the inverse dS temperature parameter. In the context of dS space, the connection between the above and the particle production as perceived by two distinct observers was studied to great extent \cite{Lapedes1978} \cite{Mottola1985} \cite{Higuchi1987}. There, a static observer placed near their event horizon and measuring the response of a field to the Bunch Davies (BD) vacuum, would experience this Hawking effect.

The BD vaccum, which was originally considered  by Chernikov and Tagirov \cite{Tagirov1968}, is defined as the state invariant under the full $SO(1,4)$ group. Hence, its corresponding modes obtain a plane-wave form at past infinity, thus provide a comfortable environment in order to calculate vacuum expectation values (VEV). As a result, in \cite{bunch1978} the two-point function and the renormalized VEV of the Energy-momentum tensor (EMT) was computed for a co-moving massive free scalar field via the point-splitting regularization technique. Even though, globally the BD state is pure, inside their causal patch, a static observer would measure a thermal mixed state caused by the tracing out of the inaccessible region \cite{Strominger2001}. Contrastingly, for a co-moving observer it retains its ``pureness", is seemingly empty as long as the initial mode basis is preserved however, a non-zero particle number
\bad
 N_\tmk = {\rm Tr}\left( \rho_{\rm out}  a^+ _{\tmk;\rm out}  a^- _{\tmk;\rm out}\right)
\ead
is found when considering an alternative `out' basis. Here, $ a^\pm _{\tmk;\rm out}$ are the ladder operators which define the new physical vacuum state and $\rho_{\rm out}$ the corresponding density matrix. While the thermal nature of the BD vacuum is manifest in the static observer frame, the same vacuum should still contain implicit information about these effects for the comoving observer. Hence, \cite{Mottola2014, Mottola2014_2} considered the case of a thermal-mixed state in the global coordinates of dS along with a time-dependent BT which resulted into the calculation of the VEV of the EMT. Through a semi-classical backreaction analysis, they demonstrated that these perturbations can become sufficiently large and alter the dS background.

%
%Most recently \cite{Chopping2024} considered the late-time cosmological correlators defined by states built from a BT of the BD vacuum, and discussed how they satisfy the conformal Ward identities of the boundary theory.
%

The main motivation of \cite{FotisAntonis3} was to examine whether similar effects arise in the expanding patch for a scalar test-field and thus how does its thermal backreaction affect the dS geometry. Towards that purpose, the Thermofield dynamics (TFD) formalism \cite{Takahashi} was used in order to rotate the BD vacuum to its thermal analog \footnote{For a connection between the TFD and the closed-time contours in the context of dS we refer to \cite{FotisAntonis2}.}. This approach provided the technical advantage of treating the thermal state as pure and thus enabled the treatment of the thermal average of the EMT as a simple VEV (see e.g. \cite{Witten2022}, for a recent discussion of TFD and the notion of entropy in dS). In the dS temperature case, it was shown that the introduced thermal corrections absorb the zero-temperature result so that the residue corresponds to the extracted thermal effects from the BD vacuum. The latter in the Poincare patch admit two possible interpretations: either they correspond to an explicit thermal bath, or describe an intrinsic thermodynamic behavior of the system, effectively characterized by $\beta_{\rm dS}$. In this interpretation, the observed thermal behavior originates from horizon-induced coarse-graining over inaccessible degrees of freedom, as modes cross the dS horizon. 
The thermal EMT was then computed via the point-splitting regulator and was plugged in the RHS of the Einstein equations. At first order in Planck mass $M_{\rm pl} = (8 \pi G)^{-1/2} $, a deviation to the original dS metric was computed with its physical interpretation being an increase/decrease to the spacetime expansion, depending on the test-field parameters.

In this paper, we take advantage of the new metric that retains its FLRW form, interpret it as a quasi–dS spacetime, and propose two distinct physical applications. The starting point is the scalar equation of motion (eom), which in conformal-time coordinates acquires a Whittaker form rather than the usual Bessel form. We then quantize the scalar field and discuss its normalization condition, which can be viewed as an extension of its analogue in de Sitter spacetime. By requiring the solution to be asymptotically equivalent to the Bunch–Davies vacuum towards the future boundary, we establish a non-trivial connection between the two bases and thereby set up the framework on which the remainder of the paper is built.

The first application focuses on an effective inflaton description sourcing the quasi–dS background and resulting in a scale-dependent curvature power spectrum. This framework is physically meaningful only in a specific regime that includes the case of a massless minimally coupled test field. In particular, we obtain a blue-tilted scalar spectrum characterized by a spectral index $n_S \sim 2$. We argue that the resulting spectrum arises only for modes exiting the horizon during a later phase of inflation and should therefore not be identified with the nearly scale-invariant spectrum observed on CMB scales. This suggests that thermal backreaction could play a role in facilitating a transition between the slow-roll regime and a transient phase relevant for primordial black hole (PBH) formation \cite{Motohashi2020, Kristiano2022}. For a recent review on observational evidence of PBH formation we refer to \cite{Carr2023}. At leading order, the resulting dynamics resemble the family of constant-roll models first studied in \cite{Starobinsky2018}.

The second application focuses on the dS/CFT holographic proposal which, unlike its AdS counterpart, involves several subtleties. During our analysis, we remain within the complementary series so that the field does not oscillate outside the event horizon and the main non-unitary issues of the dual CFTs can be avoided. Then, by assuming that the Hartle-Hawking wavefunctional is the CFT generating functional, as considered in \cite{Maldacena1}, we find in the standard quantization, the coefficient of the conformal two-point function following the holographic renormalization procedure first developed in \cite{Skenderis2002}. In \cite{Skenderis2024} the latter was further developed in the context of dS/CFT using the in-in formalism. This coefficient encodes the bulk information even though the background recovers its pure dS form at the future boundary. Away from the future boundary, which we interpret as a UV fixed point (FP) of a 3-dimensional QFT, time-evolution is treated as an inverse scale transformation. Following the Hamiltonian-Jacobi formalism and the Wilsonian picture in holography \cite{Polchinski2011, Liu2011}, we then construct the flow equations governing this dual QFT. We argue that tracing out UV degrees of freedom in the dual theory has a direct correspondence with the coarse-graining of bulk degrees of freedom, thereby motivating the use of the TFD framework in the bulk description. Our results further support the proposed duality between dS and $O(N)$ or $Sp(N)$ models \cite{Anninos2013, Das2013}, including arbitrary conformal weights within the complementary series.

The paper is organized as follows. In \sect{Quant}, we solve the mode-function equation of motion induced by the deformed background and discuss its normalization condition. In \sect{power-spectrum}, we solve the corresponding Mukhanov–Sasaki equation in the comoving gauge and compute the scalar power spectrum in the superhorizon limit, followed by a discussion of the physical interpretation of the resulting blue tilt and its connection to constant-roll models. Section \ref{Holography} is divided into two parts: the first focuses on the dual description at future infinity, while the second examines the RG flow away from it. For completeness, the appendices collect useful relations for Whittaker functions, along with brief reviews of the constant-roll models and the Wilsonian Hamilton–Jacobi formalism relevant to this work. We work in natural units $\hbar = c = 1$ and use the metric signature $\{-,+,+,+\}$.

\section{Canonical quantization in quasi-dS} \label{Quant}

\subsection{The solution to the bulk eom}
Our starting point is the toy model of a light scalar field $\phi$ propagating freely in a 4d quasi-dS spacetime which is described via a metric of FLRW form: 
\bad
ds^2 = - dt^2 + a^2(t) dx^2 = a^2(\tau) \left(- d\t^2 + dx^2 \right) 
\ead
where we denote $\t$ as the conformal time. The corresponding action is
\bad \label{Action}
{\cal S} = \frac{1}{2} \int d \t d^3 x \sqrt{-g} \left\{g^{\m \n} \partial_\m \phi \partial_\n \phi + \m^2 \phi^2 \right \}
\ead
with $\mu^2 = m^2 + \xi R$.  Here, $R = \frac{6 \ a ''(\t)}{a^3(\t)}$ is the scalar curvature with the prime denoting differentiation with respect to $\t$, while in the dS case the former is equal to $12 H^2_0$ with $H(\t) = \frac{a'(\t)}{a^2(t)}$ defined generally as the Hubble parameter. At late times ($H t \rightarrow \infty$), we assume that the scale factor $a(t)$ is written as a first order correction to the corresponding dS scale factor:
\bad \label{Quasi-dS a(t) 1}
a(t) = a_{0} (t) \left[1 + {\cal C}(m,\xi)  \frac{H^2_{0}}{M^2_{\rm pl}} e^{- H_{0} t}  + {\cal O} \left(\frac{H^4_0}{M^4_{\rm pl}}\right) \right], 
\ead   
with ${\cal C}(m,\xi)$ a real function of the mass $m$ and coupling $\xi$ to gravity of the scalar field. Then, the definition of $dt = a d \t$ lets us express conformal time   
\bad \label{correct tau}
\t = \int d \t = \int \frac{dt}{a(\t)} \simeq \int dt \ e^{- H t} \left(1 + {\cal C}(m,\xi)  \frac{H^2_{0}}{M^2_{\rm pl}} e^{- H_{0} t} \right)  = \t_{0} \left(1 + \frac{1}{2} {\cal C}(m,\xi)  \frac{H^2_{0}}{M^2_{\rm pl}} e^{- H_{0} t}  \right)
\ead 
in terms of $t$ and $\t_{0} = - \frac{1}{H} e^{- H_{0} t} $ the dS conformal time. Since the correction term is next to leading order and drops exponentially at late times, we can assume $\t \simeq \t_{0}$ and the scale factor in \eq{Quasi-dS a(t) 1} can be read as
\bad \label{Quasi-dS a(t)}
a(t) = a_{0} (t) \left[1 + {\cal C}(m,\xi)  \frac{H^3_{0}}{M^2_{\rm pl}}  |\t|  + {\cal O} \left( \frac{H^4_{0}}{M^4_{\rm pl}}\right) \right] \ .
\ead     
We note that this is a usual assumption made when one considers quasi-dS spacetimes. For instance, in slow-roll the corresponding scale factor ansatz is $a(\t) = - ( H \t_{0})^{-1}$ with $H$ a function of time which is assumed to be nearly constant. 

A scale factor of the form \eqref{Quasi-dS a(t)}, was found in \cite{FotisAntonis3} where a semi-classical backreaction to the original dS spacetime was considered due to thermal-like fluctuations. It was suggested that the latter arise when a comoving observer integrates out the lost information due to horizon crossing so that near the end, the system can be regarded as a thermal equilibrium described by the inverse Gibbons-Hawking temperature $\beta_{\rm dS} = 2 \pi/ H$. From the static's observers perspective this corresponds to the Unruh effect. The resulting correction was found to be proportional to
\bad \label{Ising4corr}
{\cal C} (m, \xi) = - \frac{1}{144 \pi^3} \nu \cot ( \pi \nu) ( 1 + 12 \eta^{-1} - 6 \xi) , 
\ead
with
\bad
\nu = \sqrt{\frac{9}{4} - \frac{\mu^2}{H^2_0}}
\ead
and $\eta = R/ m^2$. In order to not lose any generality, in this section and thereafter we will consider a general form for ${\cal C} (m, \xi)$ and explicitly note when we substitute \eq{Ising4corr} which from now on, we denote as ${\cal C}_\beta  (m, \xi)$.

Moving on, up to boundary terms the scalar field satisfies on-shell the e.o.m.:
\bad
\left( \Box  - \mu^2 \right) \phi = 0 
\ead
and due to the fact that the metric has an FRW from, the above can be massaged into
\bad
\phi '' - \nabla^2 \phi + 2 \frac{a'}{a} \phi' + a^2 \mu^2 \phi = 0 \ . 
\ead
Naturally, we consider the 3d-Fourier transform $\phi_\tmk (\t)$ of the field:
\bad \label{quadi-eom-position-space}
\phi(\t, \tx) = \frac{1}{(2 \pi)^{3/2}}\int d^3 k \ \phi_\tmk (\t) \ e^{i \tk \tx}
\ead 
which satisfies
\bad \label{Quasi-eom}
\phi''_\tmk  + 2 \frac{a'}{a} \phi'_\tmk + \left( \tmk^2 + a^2 \mu^2 \right) \phi_\tmk = 0 \ . 
\ead
The goal of this section is to solve \eq{Quasi-eom} and build a basis defined from its two linear independent solutions $u_\tmk, u^*_\tmk$, which in return will us lead to the canonical quantization for our case. This will prove to be crucial moving forward, when we compute the curvature power spectrum and the response of the boundary theory with respect to the bulk. 

We start by making the field redefinition:
\bad \label{modef}
f_\tmk (\tau) =  a (\tau) \  \phi_\tmk (\tau)
\ead
and write \eq{Quasi-eom} in terms of the $f_\tmk$ variable. We obtain
\bad \label{Quasi-eom-f}
 f''_\tmk + \left(\tk^2 +  a^2 \mu^2  - \frac{ a''}{ a } \right) f_\tmk = 0 \ .
\ead
Then, by substituting the scale factor \eqref{Quasi-dS a(t)} and expanding up to first order in $\frac{H^2_0}{M^2_{\rm pl}}$, we reach the ODE:
\bad
 f''_\tmk + \left(\tk^2 - \frac{2 \left(\frac{\mu^2}{H^2_0} +1 \right) \frac{H^3_0}{M^2_{\rm pl}} {\cal C} (m,\xi)}{\tau} + \frac{\frac{1}{4} - \nu^2}{\tau^2} \right) f_\tmk = 0 \ . 
\ead
Notice that compared to the dS case, the above equation contains an extra term that is subdominant at late-times and vanishes when ${\cal C} (m,\xi) = 0$, as expected. To clarify things even further, we define the imaginary variable $z = 2 i \tmk \tau$ for $\tmk \neq 0$ and rewrite the above eq. accordingly:
\begin{subequations}
\bad \label{Whittaker}
\frac{d^2}{dz^2} f_\tmk (z) + \left( - \frac{1}{4} + \frac{\kappa}{z} + \frac{\frac{1}{4} -  \nu^2}{z^2}\right) f_\tmk (z) = 0
\ead
with
\bad \label{Whittaker_kappa}
\kappa = i \left(\frac{\mu^2}{H^2} + 1 \right)  \frac{H^3_0}{M^2_{\rm pl} \tmk} {\cal C} (m,\xi)  \ .
\ead
\end{subequations}
The differential \eq{Whittaker} is of a Whittaker form, whose general solution is the linear combination \cite{Abramowitz}:
\bad \label{Whittaker_gen_sol}
f_\tmk (z) = A_\tmk \ M_{\kappa,  \nu} (z) + B_\tmk \ W_{\kappa, \nu} (z)
\ead
where the functions $M_{\kappa, \tilde \nu} (z), W_{\kappa, \tilde \nu} (z)$ are the Whittaker functions of first and second kind respectively as given by \eq{Whittaker_sol_M} and \eqref{Whittaker_sol_W}. In the recent past, there have been works in quasi-dS spacetimes with either different physical motivation that lead to a similar eom \cite{Frob2014,Eckardstein2024}, or slightly different $a(\t)$ that also calculate corrections to the power spectrum \cite{Chen2015}. For example, the former focus on the Schwinger effect in either scalar QED or axion inflation while the latter start with a scalar field and a perturbed metric. In both cases, their respective solutions obtain the Whittaker form with respect to the imaginary variable $2 i \tmk \t$ but with different parameters which in the latter case are completely fixed e.g. $\kappa =0$. Regarding the Schwinger effect, the number density of particle-pair production  and the VEV of the induced current have already been calculated in \cite{Frob2014}. Additionally, the similarities between the backreaction process due to particle production in dS spacetimes and in a constant electric field have already been pointed out in \cite{Mottola2014}, \cite{Mottola2014_2}. Hence, this analysis provides another interesting physical scenario for inflation when ${\cal C}(m,\xi) = {\cal C}_\b (m,\xi)$ while also extending the necessary mathematical framework for obtaining the power spectrum from generalized Whittaker functions.

\subsection{Normalization Conditions} \label{Norm_Cond}

The complex coefficients $A_\tmk, B_\tmk$ in \eq{Whittaker_gen_sol}, can be constrained via the normalization condition that arises, as usual, from the quantization of the system. In particular, if we define the canonical momentum of $\phi(\t,\tx)$ as $\pi (\t , \tx) \equiv \frac{\partial \sqrt{-g} {\cal L}}{\partial \phi'} = -a^2 (\t) \phi'$ and Fourier expand $\phi(\t , \tx)$:
\bad
\phi (\tau , x) = \frac{1}{(2 \pi)^{3/2}}\int d^3 \bm k  \ \left[\a^-_\tk u^*_\tmk (\tau)  + \a^+_{-\tk} u_\tmk (\tau) \right]e^{ i \tk \tx} 
\ead
the equal-time commutation relation
\bad
\left[ \phi(\t,\tx) , \pi (\t , \ty)\right] = i   \delta^{3} (\tx - \ty)
\ead
leads in momentum space to the conditions
\begin{subequations}
\bad
\left[\a^-_\tk, \a^+_\tq \right] = \delta^{(3)} (\tk  - \tq), \qquad \left[\a^-_\tk, \a^-_\tq \right] = \left[\a^+_\tk, \a^+_\tq \right] = 0,
\ead
\bad \label{NomrCondu}
 a^{2} (\t) \ W \left(u_\tmk, u^*_\tmk \right) = -i  \ .
\ead
\end{subequations}
Here $W(f,g) = f g' - f' g$ is the Wronskian \footnote{Notice the opposite sign compared to the usual convention starting from \eq{Action}.} and the mode-functions $u_\tmk,u^*_\tmk$ complete a set of two linear independent solutions to \eq{Quasi-eom} which are also complex conjugate to each other, so that $\phi(\t,\tx)$ is a Hermitian operator.

After the redefinition \eqref{modef}, the normalization condition \eqref{NomrCondu} corresponds to
\bad \label{Wronsk_chi}
W \left(\chi_\tmk, \chi^*_\tmk \right) = -i
\ead
where $\chi_\tmk  (\t) = a(\t) \ u_\tmk (\t)$ and its complex conjugate should form a basis of solutions to \eq{Quasi-eom-f}. Mathematically, this is permitted since \eq{Quasi-eom-f} has real coefficients and the derivatives are with respect to the real variable $\t$, which means that if $\chi_\tmk$ is a solution, then $\chi^*_\tmk$ is as well. In addition, Abel's identity \cite{DLMF} (1.13.5) restricts the Wronskian of any two linear-independent solutions to a constant \footnote{This is due to the fact that \eq{Quasi-eom-f} does not contain any first derivatives.}. Thus, we expect that the normalization condition \eqref{NomrCondu} will lead to a formula relating the complex coefficients $A_\tmk, B_\tmk$ coefficients as in the usual dS case. When writing $\chi_\tmk$ as a linear combination of the Whittaker functions, \eq{Wronsk_chi} leads to 
\bad \label{NormCondAB}
|A_\tmk|^2 W \left[M_{\kappa, \nu}\ , \left(M_{\kappa, \nu} \right)^*\right]& +2i {\rm Im} \left\{ A_\tmk B^*_\tmk W \left[M_{\kappa,  \nu}\ , \left(W_{\kappa,  \nu} \right)^*\right] \right\}  +  |B_\tmk|^2 W \left[W_{\kappa,  \nu}\ , \left(W_{\kappa,  \nu} \right)^*\right]  = -i \ . 
\ead 
This means that we only need to calculate the three Wronskians of the Whittaker functions with respect to $\t$ which can be related to their respective ones with respect to $z$ via the use of the chain rule. For example, the first Wronskian can be written as
\bad
W \left[M_{\kappa, \nu}\ , \left(M_{\kappa,  \nu} \right)^*\right] =  2 i \tmk \tilde W \left[M_{\kappa, \nu}\ , \left(M_{\kappa,  \nu} \right)^*\right]
\ead
and similar for the other two. 

Furthermore, a careful analysis of the complex conjugate of the Whittaker functions is required for our case, which we now carry out by considering each term of \eqref{Whittaker_sol_M} and \eqref{Whittaker_sol_W} separately. Recall that $z = 2 i \tmk \tau$ was defined as a pure imaginary variable located in the negative imaginary axis (since $\t<0$). 
Assuming a light field $\phi(\t,\tx)$ i.e. $\mu^2< 9H^2/4$ so that $\nu \in \mathbb R$, the power law term with the real exponent $ \n + \frac{1}{2}$ introduces a branch cut at the negative real axis. We write in the principal branch
\bad
z^{ \n + \frac{1}{2}} = e^{( \n + \frac{1}{2}) \ln z}, \quad \arg z \in ( - \pi , \pi]
\ead
where $\arg z = - \pi/2$. In this branch, the complex conjugate satisfies
\bad
\left( (2 i \tmk \t)^{ \n + \frac{1}{2}} \right)^*   =  (-2 i \tmk \t)^{\n + \frac{1}{2}} \ . 
\ead
Note that the result above can be interpreted as an analytic continuation in the complex plane along a counter-clockwise path of angle $ \Delta \theta = \pi$. This rotation moves the argument from $-\frac{\pi}{2}$ to $\frac{\pi}{2}$ reaching $z^*$ while remaining in the principal branch. Next, using the expansion of the first confluent hypergeometric function \eqref{Kummer_sol}, the complex conjugate of the Pochhammer symbol $(a)_n$ is simply $(a^*)_n$ and $( z^n)^* = (z^*)^n$ since $n \in \mathbb Z$. Hence, $\left( M(a,b,z) \right)^* = M(a^*,b,z^*)$ and recall that the Whittaker parameter $\kappa$ via \eqref{Whittaker_kappa} is imaginary leading to:
\bad
\left( M \left(\frac{1}{2} +  \nu -  \kappa, 1 + 2  \nu, z\right) \right)^* = M \left(\frac{1}{2} +  \nu +  \kappa, 1 + 2 \nu, -z\right) \ . 
\ead
Similarly, the second confluent hypergeometric functions also satisfies $\left( U(a,b,z) \right)^* = U(a^*,b,z^*)$
which can be understood via its integral representation \eqref{Tricomi_sol}. As a result, we have showed that the complex conjugates of the two Whittaker functions for $\kappa,Z \in \mathbb I$ satisfy
\bad
\left( M_{\kappa, \nu} (z) \right)^* = M_{-\kappa, \nu} (-z), \qquad \left( W_{\kappa,  \nu} (z) \right)^* = W_{-\kappa, \nu} (-z) 
\ead
which enables us to calculate each Wronskian in \eqref{NormCondAB}.

Starting with 
\bad
\tilde W \left[M_{\kappa,  \nu} (z)\ , \left(M_{\kappa,  \nu} (z) \right)^*\right] = \tilde W \left[M_{\kappa,  \nu} (z)\ , M_{-\kappa, \nu} (-z)\right]
\ead
from the analytic continuation formula \eqref{Analytic_cont M}, we can see that $M_{\kappa,  \nu} (z)\ , M_{-\kappa, \nu} (-z)$ are linear dependent and as a result their Wronskian:
\bad
\tilde W \left[M_{\kappa,  \nu} (z)\ , M_{-\kappa, \nu} (-z)\right]  = 0 \ . 
\ead
For the other two Wronskians on the RHS of \eq{NormCondAB} we can use the known results \eqref{Whittaker_Wronsk} so that \eq{NormCondAB} results into:
\bad \label{NormCondAB2}
 |B_\tmk|^2 e^{- \kappa \pi i} + 2 \ {\rm Im} \left\{   A_\tmk B^*_\tmk \frac{\Gamma (1 + 2 \nu)}{\Gamma \left(\frac{1}{2} +  \n + \kappa \right)} e^{- i \pi  \n  } \right\}    = -\frac{1}{2 \tmk} \ . 
\ead
As a consistency check, we can consider the limiting case of $\kappa  = 0$ which transfers us back to dS space, so that the differential \eq{Whittaker} can be massaged into the Bessel equation and its general solution being the linear combination
\bad \label{Bessel_gen_sol}
f_\tmk(z) = \sqrt{\tmk |\t |} \left(\tilde A_\tmk J_\nu (\tmk |\t |) + \tilde B_\tmk Y_\nu (\tmk |\t |) \right) 
\ead
of the standard Bessel functions $J_\nu (\tmk |\t|), Y_\nu (\tmk |\t|)$ while the expansion coefficients satisfy the normalization condition
\bad \label{BesselNormCondAB}
\tilde A_\tmk \tilde B^*_\tmk - \tilde A^*_\tmk \tilde B_\tmk = \frac{i \pi} {2 \tmk} \ . 
\ead 
Consequently, we should be able to reproduce the above formula from \eq{NormCondAB2} for $\kappa =0$. Legendre's duplication formula for the Gamma functions let us write
\bad
2^{2\nu} \Gamma ( \n + 1 )   = {\sqrt{\pi}} \frac{\Gamma(1+2 \nu)}{\Gamma (\n + \frac{1}{2})} 
\ead 
and with the connection formulae \eqref{Whittaker_Bessel}, we can associate the Whittaker functions when $\kappa =0$ with the modified Bessel functions $K_\n (z), I_\n (z)$. Then, the general solution \eq{Whittaker_gen_sol} can be expressed as
\bad \label{Whittaker_gen_sol_kappa=0}
f_\tmk (z) = A_\tmk  \frac{\Gamma(1+2 \nu)}{\Gamma (\n + \frac{1}{2})} \sqrt{\pi z} I_\n \left( \frac{z}{2} \right) +  B_\tmk \sqrt{\frac{z}{\pi}} K_\n \left( \frac{z}{2} \right) \ . 
\ead
Moving on, we express $K_\n (z), I_\n (z)$ in terms of the Bessel functions \cite{DLMF} (10.27.8), (10.27.6):
\begin{subequations}
\bad
K_\n \left(\frac{z}{2}\right) =\frac{1}{2} \pi i e^{ i\pi \n /2} H^{(1)}_\n (\tmk |\t|) =  \frac{1}{2} \pi i e^{i \pi \n /2} \left\{J_\n (\tmk |\t|) + i Y_\n (\tmk |\t|) \right\} 
\ead
\bad
I_\n \left(\frac{z}{2}\right) = e^{- i \pi \n /2} J_\n (\tmk |\t|)
\ead
\end{subequations}
and substitute them back into \eq{Whittaker_gen_sol_kappa=0}. This results into \eq{Whittaker_gen_sol_kappa=0} with 
\begin{subequations}
\bad \label{kappa = 0 A_BT}
\tilde A_\tmk = \sqrt{2\pi} \left\{ A_\tmk  \frac{\Gamma(1+2 \nu)}{\Gamma (\n + \frac{1}{2})}  e^{- i \pi (\n + \frac{1}{2})   /2} +  i \frac{1}{2} e^{ i \pi (\n - \frac{1}{2})  /2} B_\tmk \right\} 
\ead
and
\bad \label{kappa =0 B_BT}
\tilde B_\tmk =  - \sqrt{\frac{\pi}{2}} e^{i\pi (\n - \frac{1}{2})   /2} B_\tmk \ .
\ead
\end{subequations}
Then it is straightforward to show that \eq{BesselNormCondAB} is equivalent to \eq{NormCondAB2} for $\kappa=0$, while the above connecting formulae are essentially a BT between two separate bases.

Let us return back to our discussion concerning the general solution \eq{Whittaker_gen_sol}, whose coefficients are mildly constrained from \eq{NormCondAB2}. We will now follow the common procedure for pure dS, where a physical argument such as the BD vacuum at early-times is brought up in order to completely fix $\tilde A_\tmk, \tilde B_\tmk$. The usual choice for the Whittaker case is to set $A_\tmk = 0$ \cite{Frob2014} \cite{Chen2015} \cite{Eckardstein2024} because only $W_{\kappa, \tilde \n}$ is able to match the BD asymptotic limit for large $\tmk |\t|$. In particular at early times,
\bad
W_{\kappa, \n} \sim e^{ - i \tmk \tau} \ . 
\ead
Nevertheless, since we are interested with applications at the late-time limit, our matching should be performed there as well. 

Particularly, at $\t = 0$ the scale factor \eqref{Quasi-dS a(t)} recovers its dS form and thus a natural requirement should be that the asymptotic behaviors of \eq{Whittaker_gen_sol} and \eqref{Bessel_gen_sol} are equivalent even for $\kappa \neq 0$. For $ \nu > \frac{1}{2}$, the former is equal to \eqref{Whittaker_Assympt}:
\bad \label{Latetime_Whittaker}
f_\tmk (\tmk |\t|) = &\left( A_\tmk + B_\tmk \frac{\Gamma \left(-2 \n\right)}{\Gamma\left(\frac{1}{2} - \kappa - \n \right)} \right) \ 2^{ \frac{1}{2} +  \n} e^{- i \pi \left(\frac{1}{2} +  \n \right)/2 } (\tmk |\t|)^{ \frac{1}{2} +  \n} \\ 
& \qquad + B_\tmk \left[ \frac{\Gamma ( 2  \n )}{\Gamma \left(\frac{1}{2} +  \n - \kappa \right)} 2^{ \frac{1}{2} - \n} e^{- i \pi \left(\frac{1}{2} -  \n \right)/2 } (\tmk |\t|)^{ \frac{1}{2} -  \n} + {\cal O} \left( (\tmk |\t|)^{\frac{3}{2} -  \n} \right)\right]
\ead
while the latter \cite{DLMF}:
\bad \label{Latetime_Bessel}
f_\tmk (\tmk |\t|) = \frac{1}{2^{ \n} } \left( \frac{\tilde A_\tmk}{\Gamma(1 +  \n)} - \frac{ \tilde B_\tmk \Gamma (- \n)}{\Gamma \left(\frac{1}{2} - \n \right) \Gamma \left(\frac{1}{2} +  \n \right)}\right) (\tmk |\t|)^{\frac{1}{2} +  \n} + \tilde B_\tmk \left[- \frac{2^{\n}\Gamma(\n)}{\pi} (\tmk |\t|)^{ \frac{1}{2} -  \n} + {\cal O} \left( (\tmk |\t|)^{\frac{3}{2} - \n} \right) \right]  \ . 
\ead
At exactly $\tau =0$, the surviving terms of $\phi_\tmk = a(\t) f_\tmk (z)$ can only possibly arise from the $\frac{3}{2} -  \n$ powers. Thus we introduce the matching:
\bad \label{kappa =!0 B_BT}
B_\tmk = - \tilde  B_\tmk  \sqrt{\frac{2}{\pi}} \frac{\Gamma \left(\frac{1}{2} +  \n - \kappa \right)}{\Gamma \left(\n + \frac{1}{2}\right)}  e^{-i\pi \left(  \n - \frac{1}{2} \right)/2}
\ead
where again, Legendre's duplication formula was used in order to simplify the expression. The above expression is essentially the generalization of \eq{kappa =0 B_BT} for non-zero $\kappa$. Then, we substitute \eq{kappa =!0 B_BT} into the normalization condition \eqref{NormCondAB2}, which we expand up to first order in $\kappa$. The results is: 
\bad \label{NormCondAB3}
|\tilde B_\tmk|^2 \frac{2}{\pi} \left(1 - i \kappa \pi  + {\cal O} (\kappa^2) \right) - 2 \ {\rm Im} \left\{   A_\tmk \tilde B^*_\tmk \frac{2^{2\nu+ \frac{1}{2}}}{\pi} \Gamma ( \n + 1 ) e^{-i\pi \left( \n + \frac{1}{2} \right)/2} \right\}    = -\frac{1}{2 \tmk} \ . 
\ead
Moving on, to completely constrain $A_\tmk$ and $B_\tmk$ we can take advantage of the dS coefficient $\tilde B_\tmk$ for a specific vacuum. For the BD case,
\bad \label{BDcond}
\tilde B^{\rm BD}_\tmk = -i \frac{1}{2} \sqrt{\frac{\pi}{\tmk}}
\ead
so that
\bad
|\tilde B_\tmk|^2 \frac{2}{\pi} =  \frac{1}{2 \tmk}
\ead
and \eq{NormCondAB3} reduces to
\bad \label{NormCondAB4}
 \frac{2^{2\nu}}{\sqrt{2 \pi \tmk} } \Gamma ( \n + 1 )  \ {\rm Im} \left\{  i  A_\tmk  e^{-i\pi \left(  \n + \frac{1}{2} \right)/2} \right\} = \frac{1}{2\tmk} -i \frac{\pi}{4 \tmk} \kappa  + {\cal O} (\kappa^2) \ .
\ead
As in the pure dS case, $A_\tmk \sim \tmk^{-1/2}$ at leading order. This is justified by our observation that the above condition generalizes \eq{BesselNormCondAB} for nonzero $\kappa$. One can check that at $\kappa =0$, \eq{NormCondAB4} together with the BT \eqref{kappa = 0 A_BT} reproduce the usual BD result for $\tilde A_\tmk$. 

Following, we will proceed to calculate the power-spectrum of the curvature perturbations in the superhorizon limit where the leading term will be proportional to $|B_\tmk|^2$. Hence, \eq{kappa =!0 B_BT} and \eqref{BDcond} together will enable us to relate our result with its analogue in the BD vacuum. Afterwards, the same mode function will be evaluated at the future boundary, where the leading-order term will be identified as the external source of a boundary operator and the subleading term as the response. Consequently, the explicit values of $A_\tmk, B_\tmk$ fix the coefficient of the response kernel while \eq{NormCondAB4} imposes the normalization condition to $A_\tmk$ when the BD vacuum is assumed.

\section{The enhanced curvature perturbation spectrum} \label{power-spectrum}
Now, we are ready to examine what are the consequences to the inflationary history of the universe produced by the quasi-dS spacetime described by \eq{Quasi-dS a(t)}. Firstly,  we consider the resulting time-dependent Hubble scale:
\bad \label{quasiH2}
 H = H_0 \left[1 - 2{\cal C} (m ,\xi) \frac{ H^3_0}{M^2_{\rm pl}}  |\t| + {\cal O} \left(\frac{H^4_0}{M^4_{\rm pl}} \right) \right]
\ead
from which it is clear that the linear dependence in time of the correction will produce a transient effect on the evolution of the inflationary period. This becomes even more evident if one calculates a gauge-invariant quantity like the Ricci scalar $R$ which will obtain corrections to the usual dS result. We define the Hubble flow parameters
\bad \label{epsilonH}
\epsilon_H \equiv  - \frac{H'}{a H^2} =  - 2 {\cal C} (m ,\xi)  \frac{H^3_0}{M^2_{\rm pl}} |\t| + {\cal O} \left(\frac{H^4_0}{M^4_{\rm pl}} \right)
\ead
and 
\bad
\eta_H \equiv \frac{\epsilon'}{a H \epsilon} \sim -1 \ . 
\ead
As the inflationary period approaches its ending $\t \rightarrow 0^-$, $\epsilon_H$ becomes negligible and does not match the expected behavior i.e. $\epsilon_H \rightarrow 1$. Therefore, the thermal backreaction is not a suitable candidate to describe the transition between the inflationary and reheating periods of the universe's history. However, as we will showcase below this kind of behavior can be understood as a certain case of the more-generic constant-roll models \cite{Starobinsky2018, Motohashi2020, Motohashi2025} which we briefly review in \sect{Constant-roll}.

We consider a single-field model described by the action   
\bad \label{Sifl3}
{\cal S}_{\rm ifl} = \frac{1}{2} \int d^4x \sqrt{-g} \left\{g^{\mu \nu} \partial_\mu \varphi \partial_\nu \varphi + V_{\rm eff}(\varphi)\right\}
\ead
where $\varphi$ is the minimally coupled effective inflaton producing the considered quasi-dS spacetime. Note that we have restored cosmic time coordinates since they are preferable by most cosmological conventions. In these coordinates, the field eom and Friedman eq. are given by 
\begin{subequations}
\bad \label{infleom}
\ddot \varphi + 3 H \dot \varphi + \frac{1}{2}\frac{\partial V_{\rm eff}}{\partial \varphi} = 0 \ , 
\ead
\bad \label{1stFried3}
3 H^2 = \frac{1}{ 2 M^2_{\rm pl}} \left[ (\dot \varphi)^2 + V_{\rm eff} \right] 
\ead
and
\bad \label{2ndFried3}
\dot H = - \frac{1}{ 2 M^2_{\rm pl}} ( \dot \varphi)^2
\ead
\end{subequations}
with the dot denoting the derivative with respect to cosmic time, as usual. The constant-roll proposal is centered around the assumption
\bad \label{CRcond}
\frac{\ddot \varphi}{\dot \varphi} = \alpha H
\ead 
with $\alpha$ a constant of order unity. The above condition produces a family of models characterized by $\alpha$ which is constrained between $-3<\alpha<0$ with the maxima corresponding to slow-roll and the minima to ultra-slow roll \cite{Kristiano2024, Petel2025}. In this regime, the power-spectrum obtains a blue-tilt which is why these models are of particular physical interest. In particular, they naturally generate an enhancement of the curvature power spectrum at small scales, which may lead to the formation of PBH \cite{Motohashi2020}. Such enhancements typically occur for modes exiting the horizon around 10–30 e-folds before the end of inflation. This gives enough time for the system to ``heat-up" up to $\beta = \beta_{\rm dS}$, thereby indicating that information-loss could be the origin of the spectrum enhancement. 

We substitute \eq{quasiH2} into the LHS of \eq{2ndFried3} and obtain
\bad \label{dotH}
\dot H = 2 {\cal C} (m ,\xi) \frac{ H^4_0}{M^2_{\rm pl}} e^{- H_0 t} \ . 
\ead
Then, by integrating both sides of \eq{2ndFried3} we find 
\bad \label{varphisol}
\varphi = - 4 H_0 \sqrt{- {\cal C} (m ,\xi)} e^{ - \frac{H_0 t}{2}} + {\cal C}_1
\ead
with ${\cal C}_1$ an arbitrary constant. Its specific value is not of importance, nevertheless we could for instance require that the inflaton vanishes at $t \rightarrow \infty$ so that ${\cal C}_1 = 0$. Clearly, \eq{varphisol} makes physical sense only for ${\cal C} (m,\xi) <0$. Otherwise the expectation value of the inflaton and its derivatives become imaginary resulting into a complex stress-energy tensor and a potential loss of unitarity. This restricts the parameters $m,\xi$ of the initial test-field $\phi$ probing the dS spacetime. In particular when ${\cal C} (m,\xi) = {\cal C}_\beta (m, \xi)$, this is satisfied by $1 <\nu < \frac{3}{2}$ which corresponds to a nearly massless minimally coupled test-field $\phi$.    

The above relation has two implications. Firstly, the double-dot derivative of \eq{varphisol} satisfies:
\bad \label{CRcond2}
\frac{\ddot \varphi}{\dot \varphi} = -\frac{1}{2} H_0  \sim \left(- \frac{1}{2} -  {\cal C} (m , \xi) \frac{ H_0^2}{M^2_{\rm pl}} e^{- H_0 t} \right) H
\ead
where for the last step we made use of \eq{quasiH2} in order to express $H_0 $ in terms of $H$. Then, we can directly compare \eq{CRcond} and \eqref{CRcond2} and find $\alpha \sim -\frac{1}{2}$ along with a non-leading time-dependent part of order $H^2_0 / M^2_{\rm pl}$. This places $\alpha$ in the $-3 < \alpha < 0$ regime which enables us at leading order, to read the thermal backreaction as a transient nearly constant-roll model that produces, for a few e-folds, a blue-tilted power-spectrum. In particular, we define the scalar perturbations $\delta \varphi (\t ,\tx), \delta g (\t ,\tx)$ so that
\bad
\bar \varphi (\t ,\tx) = \varphi (\t) + \delta \varphi (\t ,\tx), \qquad \bar g_{\m \n} (\t ,\tx) = g_{\m \n} (\t) + \delta g_{\m \n} (\t ,\tx)
\ead
are their total counterparts. In the co-moving gauge \cite{Maldacena1}, the inflaton remains unperturbed $\delta \varphi =0$ while $g_{\m \n}$ can be parametrized by
\bad
ds^2 = - a^2 d \t^2 + a^2 ( 1 - 2 {\cal R} ) dx^2 \ . 
\ead
with ${\cal R}$ the curvature perturbation. We define the new Mukhanov-Sasaki variable $A^2 = 2 M^2_{\rm pl} a^2 \epsilon_H$ so that: 
\bad
\frac{A''}{A} = \frac{3}{4\t^2} - 
\frac{ {\cal C} H_0^3 }{M^2_{\rm pl} |\t|}   + {\cal O} \left(\frac{H^4_0}{M^4_{\rm pl}} \right) \ . 
\ead
Substituting the above into the Mukhanov-Sasaki \eq{Mukhanov-Sassaki2} and rewriting it in terms of the variable $z = 2 i \tmk \t$ in momentum space, we arrive at the Whittaker equation
\bad
\frac{d^2 v}{d z^2} + \left( - \frac{1}{4} + \frac{\tilde \kappa}{z} - \frac{3}{4z^2} \right) v = 0
\ead
with  $v = A{\cal R} $ and 
\bad
\tilde \kappa =  i \frac{{\cal C} (m \xi)  H^3_0}{2 M^2_{\rm pl} \tmk} \ . 
\ead
Similarly to before, the general solution to the above equation is
\bad
v_\tmk = A_\tmk M_{\tilde \kappa, \n} (z) + B_\tmk W_{\tilde \kappa, \n} (z) 
\ead
with $\n = 1$ and we follow the usual quantization procedure as described in \sect{Norm_Cond}. In the super-horizon limit, we make use of the expansion \eqref{Latetime_Whittaker} whose two independent solutions give
\bad
v_\tmk = {\cal A}_\tmk (\tmk |\t|)^{3/2 } + {\cal B}_\tmk (\tmk |\t|)^{-1/2}
\ead
with ${\cal A}_\tmk, {\cal B}_\tmk$ denoting the independent coefficients of \eq{Latetime_Whittaker} for readability. The co-moving curvature mode 
\bad \label{CRR}
{\cal R}_\tmk \equiv \frac{v_\tmk}{A} \sim  {\cal B}_\tmk \tmk^{-1/2}   +  {\cal A}_\tmk \tmk^{3/2} |\t|^{2 } 
\ead
thus consists of a conserved mode and a decaying part as $\t \rightarrow 0^-$. Notice that we reach the asymptotic behavior as \eq{solutionR} for a pure constant $\alpha = - \frac{1}{2}$. Thus, we conclude that at leading order, the time dependent corrections in \eq{CRcond2} can be safely considered as negligible. This approach however, begins to differ from constant-roll at next order where we also have to take into account the second order backreaction process and the deviation from the dS conformal time \eqref{correct tau}. Therefore, a detailed analysis should find additional fluctuations contributing to the spectrum enhancement and, as a result, to the suspected PBH formation, though this exceeds the scope of this paper.

Moving on, we have argued that after the horizon exit of the CMB fluctuations, our universe entered a stage described by \eq{Quasi-dS a(t)} which at leading order has a constant-roll behavior. This expands the idea proposed in \cite{Motohashi2020} of a transient constant-roll phase over the course of ${\cal O} (10)$ number of e-folds right before the end of inflation. The modes exiting the horizon during that stage result into the power-spectrum:    
\bad \label{PR2}
 {\cal P}_{R} \equiv \frac{\tmk^3}{2 \pi^2} |R_\tmk|^2  \sim \frac{ 1}{4 
 \pi^3} \left[ -\frac{ \tmk}{{\cal C} (m,\xi) H_0} + {\cal O} \left(\frac{H^4_0}{M^4_{\rm pl}} \right) \right] 
\ead
at leading order while the residual terms become negligible in the superhorizon limit. In the above relation we made use of \eq{kappa =!0 B_BT} and the BD condition \eqref{BDcond}. Notice that the ${\cal C} (m,\xi)<0$ in the denominator signals the dependence on the test-field parameters. Specifically when ${\cal C} (m,\xi) = {\cal C}_\beta (m,\xi)$, as $\nu \rightarrow \frac{3}{2}$ the above result increases and vice-versa. This demonstrates an additional degree of freedom that is worth investigating near the massless limit. 

Additionally, the scalar spectral index is $n_S \sim 2$ which shows a strong UV tilt for the power-spectrum and is able to lead to the required ${\cal P}_{R} \sim 10^{-3} -10^{-2}$ peak for the PBH formation. In contrast, the CMB modes exhibit a nearly scale-invariant spectrum which is characterized by the experimental value $n_S \sim 0.964$ \cite{Planck1}. For a more recent re-analysis we also refer to \cite{Tristram2024}. We emphasize that $n_S \sim 2$ describes only a transient phase occurring later during inflation. During this phase, large $\tmk$ modes, corresponding to small physical scales, exit the horizon and can form PBHs due to the enhancement produced by the thermal backreaction. We argue that \eq{PR2} captures this effect and should therefore not be confused with with the slightly red-tilted spectrum observed on CMB scales.

Secondly assuming that $\varphi (\t)$ is a monotonic function of time, we can solve \eq{varphisol} in terms of $\t$ and substitute it into \eq{varphisol} into \eq{quasiH2}. This leads to
\bad \label{Hsol}
H(\varphi) = H_0 \left[1 +   \frac{\varphi^2}{8 M^2_{\rm pl}}   + {\cal O} \left(\frac{H^4_0}{M^4_{\rm pl}} \right) \right] \ . 
\ead
The above result can be read as a first order approximation of \eq{Hvarphi} for ${\cal C}_1 = H_0$:
\bad
H(\varphi) = H_0 \cosh\left[ \frac{\varphi}{2 M_{\rm pl}} \right]
\ead
which was thoroughly examined in \cite{Starobinsky2018} for general $\alpha$. In other words, by reshaping \eq{dotH} into an ODE with respect to $\varphi$, analogous to \eq{dotH}, one can find its solution via the WKB approximation and expand it around $\varphi =0$. The end result will then be \eq{Hsol}. Nevertheless, we note again that this agreement should be trusted only up to first order since for higher orders, the time-dependent part in \eq{CRcond2} cannot be neglected. Finally, we can make use of \eq{CRVeff} in order to express the effective potential:
\bad \label{Veffsol}
 V_{\rm eff} (\varphi) =  M_{\rm pl}^2 H_0^2 \left[6  + \frac{5}{4}  \frac{\varphi^2}{M^2_{\rm pl}} +{\cal O} \left( \frac{\varphi^4}{M^4_{\rm pl}} \right) \right]
\ead 
as a quadratic function of $\varphi$. We thus recover at leading order the constant-roll potential \eqref{CRVeff} for $\alpha = - \frac{1}{2}$. The quadratic potential has already been studied outside of the context of slow-roll \cite{Tzirakis2007} where a duality exists, between the different background solutions $\alpha$ and $\tilde \alpha = - (3 + \alpha)$. Specifically, one could still recover the same effect potential as above for $\alpha = - \frac{5}{2}$ yet this would lead to a non-stable growing ${\cal R}_\tmk$. Moreover, notice that at the late time limit \eq{varphisol} shows a damping of the inflaton while the potential \eqref{Veffsol} exhibits a minimum at $\varphi =0$. Simultaneously, \eq{epsilonH} gives $\epsilon_H << 1$ which hints that the effects considered here ought to remain transient and an extra mechanism is needed to describe inflation ending.

In this section we showcased that a semiclassical backreaction speculated to be produced by information loss at the late-time limit, leads to an amplification to the curvature power-spectrum. Hence, we argue that the former is a sufficient candidate to describe how inflation exits its initial slow roll phase towards an intermediate era leading to PBH formation. Nevertheless, several questions need further investigation. In particular, the exact mechanism describing the transition from the intermediate phase to the final slow-roll phase is yet to be built. 
For instance, a possible getaway would be to consider a time-dependent $H_0$ due to a slow-roll background. This will lead to the flow parameter
\bad
\epsilon_H \equiv    \epsilon_0 - 2 {\cal C} (m ,\xi) \frac{ H^3_{0}}{M^2_{\rm pl}} |\t| + {\cal O} \left(\frac{H^4_{0}}{M^4_{\rm pl}}\right)
\ead 
where we have suppressed terms proportional to $H^2_0 \epsilon_0/ M^2_{\rm pl}$. Then, $\epsilon_0$ would grow as $\t \rightarrow 0$ and dominate the extra corrections. Furthermore, this mechanism should produce an extra inflaton mass $\delta m^2 \varphi^2 \sim  \frac{5}{4} H^2_0 \varphi^2$ so that
\bad \label{Totalpotential}
 V_{\rm eff} (\varphi) =  M_{\rm pl}^2 H_0^2 \left[\text{const}  + \left(\frac{5}{4} - \frac{\delta m}{H^2_0} \right) \frac{\varphi^2}{M^2_{\rm pl}} +{\cal O} \left( \frac{\varphi^4}{M^4_{\rm pl}} \right) \right] \ . 
\ead
A similar potential was studied in \cite{Tzirakis2007} where it was shown that the slow-roll becomes a late-time attractor solution once again. Therefore, a semi-classical thermal backreation analysis for various slow-roll potentials that satisfy \eq{Totalpotential} up to first order, could possibly describe the transition towards the last inflationary phase.

%%%%%%%%%%%%%%%%%%%%%%%%%%%%%%%%%%%%%%%%%%%%%%%%%%%%%%%%%%%%%%%%%%%%%%%%%%%%

%%%%%%%%%%%%%%%%%%%%%%%%%%%%%%%%%%%%%%%%%%%%%%%%%%%%%%%%%%%%%%%%%%%%%%%%%%%%  

\section{The holographic effects} \label{Holography}

\subsection{The future infinity boundary}

The spacetime geometry considered in this paper is asymptotically dS at the late-time limit which means that at $\tau = 0$ it shares the same asymptotic boundary. Thus by assuming that the dS/CFT correspondence holds, in this section we will check how the conjectured dual theory is affected by the corrected scale factor \eq{Quasi-dS a(t)}. 

Consider the action \eqref{Action} defined in between the initial and final times $\t_i$ and $\t_f$. Integrating by parts and assuming either a vanishing field or its derivative at spatial infinity we find
\bad \label{On-Shell action}
{\cal S} = \frac{1}{2} \int^{\t = \t_f}_{\t = \t_i}  d \t d^3 \tx \ a^2(\t) \phi \left[\partial^2_\t + 2 \frac{a'(\t)}{a(\t)} \partial_\t - \nabla^2 + a^2(\t) \mu^2 \right] \phi  -  \frac{1}{2} \int d^3 \tx \  a^2 (\t) \phi \partial_\t \phi \biggl|^{\t = \t_f}_{\t = \t_i} 
\ead
with the first term vanishing on-shell due to the eom \eqref{Quasi-eom}. Additionally, in order to have a well-defined variational principal, we assume Dirichlet boundary conditions for the future (past) infinity cases where $\t_f  = 0$ $(\t_i = -\infty)$. For future reference, when $\t_f \neq 0$ instead we will assume mixed boundary conditions which will let the field to fluctuate. 
 
In the usual holographic dictionary \cite{Polyakov1998, Witten1998} adapted to the dS case \cite{Maldacena1}, the main postulate is that the wavefunctional at the future-infinity $\t_f = \epsilon$ defined by
\bad \label{Wavefunctional_e1}
\Psi [\phi(\epsilon); \epsilon] \equiv \int {\cal D} \phi \exp\left\{-  {\cal S}\right\}
\ead
can be interpreted as the generating functional 
\bad \label{CFTpart}
{\cal Z}_{\rm CFT} = \braket{e^{-  \int d^3 x J(x) {\cal O}}}_{\rm CFT}
\ead
of a CFT in Euclidean space where $J(x)$ is the external source of bulk origin corresponding to a CFT operator ${\cal O}$. The bulk field $\phi$ is decomposed into two independent parts, the leading and sub-leading from either which, we fix $J(x)$ leading to the standard or alternative quantization respectively. Here we have introduced the negative regulator with magnitude $|\epsilon| << 1$. This will let us deal with the upcoming divergences via the holographic renormalization procedure reviewed in \cite{Skenderis2002}. 

Since the eom is singular at the late-time limit, we can take advantage of Frobenius's method i.e. express the solution as a power-series of the form:
\bad \label{Frob_ansatz}
\phi_\tmk (\t) = \sum^\infty_{j=0} a_j \tmk^{\Delta+ j} |\t|^{\Delta + j} 
\ead  
with the coefficient $a_0 \neq 0$ fixed by the initial conditions of the ODE and $\Delta$ the leading order. Then, we substitute the above into \eq{Quasi-eom}, collect together similar powers of $|\t|$ and demand a trivial solution. This results into:
\bad
\Delta( \Delta - 3 ) + \frac{\m^2}{H^2}  = 0 
\ead
\bad \label{non-zero a1}
 a_1 \left[(\Delta -2)( \Delta  +1) + \frac{\m^2}{H^2_0}\right] + a_0 \frac{ 2 {\cal C} (m, \xi) H^3_0 }{M^2_{\rm pl} \tmk} \left[  \Delta + \frac{\m^2}{H^2_0} \right] = 0 
\ead
and the recurrence relation 
\bad \label{recc.rel}
a_{j+2} \left[  ( \Delta + j +2) (\Delta + j -1) + \frac{\m^2}{H^2_0}\right] + a_{j+1}  \frac{2{\cal C} (m,\xi) H^3_0 }{M^2_{\rm pl} \tmk}\left[ (\Delta + j +1) + \frac{\m^2}{H^2_0} \right] + a_j = 0 \ . 
\ead 
By solving the first equation, we find the two-independent solutions $ \Delta_\pm = \frac{3}{2} \pm \n $ which serve as the classical scaling dimensions of the two independent parts of the bulk field. From the above it is clear that our case has a similar leading behavior as dS case and by fixing either $\Delta_\pm$ results into a choice between the standard or alternative quantizations.

In the usual dS case where ${\cal C} (m ,\xi) = 0$, the second and third equations for $\Delta = \Delta_\pm$ and $2 \n \notin \mathbb Z$ force $a_1 = 0$ and $a_{2 n +1} = 0, n \in \mathbb Z$ correspondingly. Otherwise, when $\n$ is a half integer the recurrence relation breaks down and one needs to introduce logarithmic terms in the ansatz \eq{Frob_ansatz} as the usual prescription suggests \cite{Skenderis2000}. Then, the logarithmic counterterms introduce an explict scale-dependence to the on-shell action which results to the emergence of a RG flow. Notice that in \eq{Ising4corr}, $\n = \frac{1}{2}, \frac{3}{2}$ leads to ${\cal C}_\b (m ,\xi) =0$ so that dS symmetry is restored. This implies that the thermal deformation of the boundary theory and the above running do not mix. Contrastingly, an arbitrary non-zero ${\cal C} (m ,\xi)$ for $\n = \frac{1}{2}, \frac{3}{2}$ leads to a non-zero $k$-dependent $a_1$ proportional to $a_0$ and could potentially safeguard the recurrence relation. However, this case will not be examined here. Overall, note that in the light field regime  where $\frac{1}{2} \leq \n \leq \frac{3}{2}$, we obtain $\Delta_- \geq 0$ and there is no IR divergence arising from the limit $\tmk \rightarrow 0$ in \eq{Frob_ansatz}. This becomes apparent when noticing that \eq{recc.rel} results into $\tmk^{-(j+2)}$ at most for $a_{j+2}$. 

The above analysis confirms that $a_0$ remains the only free parameter of our case which we end up fixing by choosing initial conditions. In particular, we directly compare \eq{Frob_ansatz} with the asymptotic expansion of $\phi_\tmk (\t) = a^{-1} (\t) f_\tmk (\tmk |\t|)$ around $\t \rightarrow 0^-$ via \eq{Latetime_Whittaker}, and obtain the matching:
\bad
a_0 = 
\begin{cases} \label{a_0}
(-2 i  )^{\frac{1}{2} + \n} \left( A_\tmk + B_\tmk \frac{\Gamma \left(-2 \n\right)}{\Gamma\left(\frac{1}{2} - \kappa - \n \right)} \right)  \tmk^{-1} H_0, & \Delta = \Delta_+\\
 (-2 i  )^{\frac{1}{2} - \n} B_\tmk  \frac{\Gamma \left(2 \n\right)}{\Gamma\left(\frac{1}{2} - \kappa + \n \right)} \tmk^{-1} H_0 , & \Delta = \Delta_- \ . \\
\end{cases}
\ead
Let us use from now on a superscript $\pm$ to denote the value of $a_0$ for either choice and $a^\pm_j, j\neq 0$ the rest of the coefficients obtained from \eq{recc.rel}. Following the discussion of \sect{Norm_Cond}, the coefficients $A_\tmk, B_\tmk$ are constrained by the normalization condition \eq{NormCondAB2} so that we end up with one remaining degree of freedom which can be fixed by choosing the BD asymptotic form at future infinity. 

Substituting \eq{Frob_ansatz} into the Fourier transform of the on-shell action \eqref{On-Shell action} with $\t_f = \epsilon$ serving as the regulator and assuming the usual boundary conditions at past infinity, we find
\bad
{\cal S}_{\rm on-shell} = - \frac{1}{2 (2 \pi)^3} \sum^{\infty}_{j,l=0}\int d^3 \tk \ a^2(\epsilon) \ a_j^\pm a_l^\pm \ (\Delta + l) \ \tmk^{2 \Delta + j + l} \ |\epsilon|^{2 \Delta + j + l -1} \ .  
\ead 
Here $a^2(\epsilon)$ can be approximated in the semi-classical limit by expanding \eq{Quasi-dS a(t)} at first order in $H^2_0 / M^2_{\rm pl}$ and $ \Delta \in \{  \Delta_+,  \Delta_-\}$. Since most terms of the above expansion vanish at $\epsilon \rightarrow 0$, we will keep only the  finite and divergent terms in our expressions. Thus,
\bad
{\cal S}_{\rm on-shell} = {\cal S}_{\rm fin} + {\cal S}_{\rm div}
\ead
with
\bad \label{finite action}
{\cal S}_{\rm fin} =- \frac{3}{2 (2 \pi)^3} \int d^3 \tk  \ a^-_0 a^+_0 H^{-2}_0 \tmk^{3}  
\ead
and
\bad
{\cal S}_{\rm div} = - \frac{1}{2 (2 \pi)^3}  \int d^3 \tk \ & H^{-2}_0 \biggl\{   \sum_{j,l=0}^{l+j \leq 2 \n} a^-_j a^-_l \left(\Delta_- + l\right) \tmk^{2 \Delta_- + j + l} |\epsilon|^{-2 \n + j + l } \\ 
& \qquad + \frac{2 {\cal C} (m,\xi) H^3_0 }{M^2_{\rm pl}\tmk} \sum_{j,l=0}^{j+l \leq 2 \n -1} a^-_j a^-_l \left(\Delta_- + l\right) \tmk^{2 \Delta_- + j + l+1} |\epsilon|^{1-2 \n + j + l } \biggr\} \ . 
\ead 
In the usual dS case, recall that $a_{2n+1} =0$ for $n$ a positive integer so that the non-zero terms correspond to even $j,l$. Hence, it becomes apparent that the asymptotic dS spacetime considered in this work, requires the introduction of extra counterterms in order to cancel the additional divergences. 

The leading and next to leading order divergences proportional to $|\epsilon|^{- 2 \n}, |\epsilon|^{1- 2 \n}$ are canceled entirely by the 1st-counterterm 
\bad \label{1st counterterm}
{\cal S}^{(1)}_{\rm ct} \equiv \frac{1}{2 }  \int d^3 \tx \sqrt{-\gamma} c_1 \phi^2 = \frac{1}{2 (2 \pi)^3}  \int d^3 \tk \sqrt{-\gamma} c_1 \phi_\tk \phi_{-\tk} 
\ead 
which we wrote directly in momentum space where $\gamma_{ij} = a^2(\t) \delta_{ij}$ is the induced metric on the boundary. Using \eq{Frob_ansatz}, the coupling $c_1$ obtains the form
\bad 
c_1 = c^{(0)}_1 + c^{(1)}_1 |\epsilon|
\ead
with
\bad \label{c_1}
c^{(0)}_1 = \left(\frac{3}{2} - \n \right) H_0
\ead
the $\epsilon$-independent part and
\bad
c_1^{(1)} =  -\frac{3{\cal C} (m,\xi) H^4_0}{M^2_{\rm pl}} \frac{3 - 2\n }{1 - 2 \n} 
\ead 
the coefficient of the correction. Evidently, when $\n = \frac{3}{2}$ or ${\cal C} (m,\xi) = 0$ in general, the coupling is restored to its pure-dS value \cite{Skenderis2002}, as it should. This suggests that even for the case where \eq{Ising4corr} does not hold i.e. ${\cal C} \neq {\cal C}_\b$, the massless minimal coupled case (or more generally $\m=0$) introduces the same counterterms as the dS case. Otherwise, the persisting $\epsilon$-dependent part of $c_1$ is expected via the finite terms of \eq{1st counterterm} to introduce couplings into the total action which can be interpreted as scale-dependent. We expect that these will alter the emergent RG flow as $\epsilon$ moves away from zero, as a result. 

In the $1 < \n < \frac{3}{2}$ regime, an extra counterterm needs to be considered in order to cancel the $|\epsilon|^{2- 2 \n}$ divergences. Thus, we introduce:
\bad \label{2nd counterterm}
{\cal S}^{(2)}_{\rm ct} \equiv - \frac{1}{2 }  \int d^3 \tx \sqrt{-\gamma} c_2 \phi (x) \Box_\gamma \phi (x)  = \frac{1}{2 (2 \pi)^3} \int d^3 \tk \  a(\epsilon)  \tmk^2 c_2 \phi_\tk \phi_{-\tk}
\ead 
with $\Box_\gamma$ the D'Alambertian operator defined by the induced metric. Then, the leading order solution to $c_2$ is
\bad
c_2 = - \frac{1}{2 (1- \n)} H^{-1}_0 + {\cal O} \left(\frac{H^4_0}{M^4_{\rm pl}}\right) \ , 
\ead      
which agrees with the dS result. At this point, we should note that the second order terms have a non-trivial $\tmk$-dependence which we choose to omit, since we have already truncated all our previous results at first order. 

The renormalized action is then defined as
\bad \label{Sren}
{\cal S}_{\rm ren} = {\cal S} + {\cal S}_{\rm ct}
\ead
so that only the finite part of \eq{1st counterterm} and \eqref{finite action} remain in the $\epsilon \rightarrow 0$ limit:
\bad
\lim_{\epsilon \rightarrow 0} {\cal S}_{\rm ren} = - \frac{1}{ (2 \pi)^3} \int d^3 \tk  \ \n a^-_0 a^+_0 H^{-2}_0  \tmk^{3}  \ . 
\ead
As discussed earlier, in the standard quantization we read the $|\epsilon|^{\Delta_-}$ contribution of $\phi_\tmk$ as the source of a dual operator ${\cal O}$. If we denote $\phi_\tmk \sim \phi^-_\tmk |\epsilon|^{\Delta_-} + ... $ from \eq{Frob_ansatz} we find 
\bad \label{Sren}
\lim_{\epsilon \rightarrow 0} {\cal S}_{\rm ren} =  - \frac{1}{ (2 \pi)^3} \int d^3 \tk \ \phi^-_{\tk} G \phi^-_{-\tk},
\ead 
with
\bad \label{Gcft}
G = \n (-2i)^{2\n} \frac{1}{\Gamma(2\n)} \left(\frac{A_\tmk}{B_\tmk} + \frac{\Gamma(-2\n)}{\Gamma \left( \frac{1}{2} - \kappa - \nu \right)} \right) H^{-2}_0 \tmk^{2 \n} \  .
\ead
In the above, we started from \eq{a_0} which enabled us to express $G$ in terms of the coefficients $A_\tmk, B_\tmk$ of the general solution \eqref{Whittaker_gen_sol}. Hence, \eq{Gcft} can be understood as the extension of its pure dS analogue corresponding to $\kappa \neq 0$.  As consistency check, we verify that the above relation reduces to the de Sitter case when we require that $\phi_\tmk$ behaves like a pure dS field near $\t \rightarrow 0$. For this purpose, we use \eq{kappa =!0 B_BT} and match the non-leading terms of \eq{Latetime_Whittaker} and \eqref{Latetime_Bessel} in order to express the ratio $A_\tmk/B_\tmk$ in terms of $ \tilde A_\tmk/\tilde B_\tmk$. Furthermore, we also make use of the BD vacuum which sets $\tilde B_\tmk \propto -i \tilde A_\tmk$ so that \eq{Bessel_gen_sol} obtains a Hankel function form. Then, \eq{Gcft} has the form of a CFT 2pt function $
\braket{\cal {\cal O}_\tmk {\cal O}_{-\tmk}} = {\cal C_O} \tmk^{2 \Delta_+ -3}$ and the normalization constant is
\bad
{\cal C_O} =  2^{-2\n}  \pi \n (-i)^{2 \n-1} \frac{1}{\Gamma(\n)} \left(\frac{1}{\Gamma (1 + \n)}i + \frac{\Gamma (-\n)}{\Gamma \left(\frac{1}{2} - \n \right) \Gamma \left( \frac{1}{2} + \n \right)} \right) \ . 
\ead 
In position space this corresponds to \cite{Skenderis2013}:
\bad
\hat {\cal C}_{\cal O}  = \frac{\Gamma \left( \frac{3}{2} + \n \right)}{\pi^{3/2} 2^{1-2\n} \Gamma (-\n)}  {\cal C_O} = \frac{\n}{2 \pi^{1/2}} (-i)^{2 \n -1} \left(\frac{\Gamma \left(\frac{3}{2} + \n \right)}{\Gamma(-\n) \Gamma(\n) \Gamma(1+ \n)}i + \frac{\frac{1}{2} + \n}{\Gamma \left( \frac{1}{2} - \n \right)} \right)  
\ead
where $\hat {\cal C}_{\cal O} \equiv \braket{\cal {\cal O}(\tx) {\cal O}(\ty)} |\tx-\ty|^{2 \Delta_+}$. Notably, we have arrived at a CFT propagator of an arbitrary operator ${\cal O}$ of scaling dimension $\Delta_+$ and this serves as clear indication for the dS/CFT conjecture. Particularly, in the saddle-point approximation of \eq{Wavefunctional_e1}, \eq{Sren} leads to:
\bad
\Psi [\phi(\epsilon); \epsilon] = \exp\left\{  - \frac{1}{(2\pi)^3} \int d^3 x d^3 y  \  J(\tx) G(\tx-\ty) J(\ty) \right\}
\ead
where the RHS has the form of a CFT partition function coming from \eq{CFTpart} with $J(x) = H^{-1}_0 \phi^-(x)$ acting as its external source. 
On the other hand, the alternate quantization requires us to read the coefficient $\tilde \varphi_\tmk$ of the subleading solution $\phi_\tmk \sim \phi^+_\tmk |\epsilon|^{\Delta_+}$ as the latter. Consequently, the boundary operator ${\cal O}(x)$ will be of dimension $\Delta_-$ and the CFT unitarity bound requires $\n \leq 1$. For this reason, the above analysis focuses on the standard approach in order to keep our treatment as general as possible.

Although, the CFT symmetry is restored at the future boundary due to the vanishing of the intermediate corrections in \eqref{Quasi-dS a(t)}, the existence of the latter for $\t \neq 0$ leads to initial conditions that depend on $\kappa$. As a result, the modified subleading behavior of the field compared to the dS case gives rise to the more general form of \eq{Gcft}. This can also be understood by $G(x-y)$, which should be interpreted as the boundary response kernel to the external source $J(x)$ originating in the bulk, and therefore naturally encode the corresponding  bulk information.

\subsection{The Wilsonian Holographic RG flow}
As we explored in the previous subsection, the future boundary $\t =0$ of our 4d spacetime corresponds to a fixed point of a 3d dual theory containing an operator ${\cal O}$ whose scaling dimension is $\Delta_+ (\Delta_-)$ in the standard (alternate) quantization. Clearly, as we move away from the boundary, the bulk effects lead to the emergence of finite $\epsilon$-dependent terms in \eq{1st counterterm} which in return will lead to a deformed RG flow of the boundary theory. 

In this subsection, we will consider how the dual theory is probed as we push away from $\t =0$ via a backwards time-evolution which we will effectively treat as an inverse scale transformation. Towards that purpose, we will mainly follow \cite{Polchinski2011, Liu2011} which extend the Wilsonian picture for a holographic setup where one integrates out UV degrees of freedom. For earlier attempts to build this analogy we refer to \cite{deBoer1999, Larsen2} for the AdS and dS cases respectively. We will first provide a quick review of the formalism in order to define the main quantities and then consider our specific case. 

Let us assume a cut-off at $\t = \t_c \neq 0$ so that the renormalized action \eqref{Sren} can be decomposed as
\bad
{\cal S}_{\rm ren} = \int^{\t_c}_{-\infty} d \t \ \sqrt{-g} {\cal L} + \int^0_{\t_c} d \t \ \sqrt{-g} {\cal L} + {\cal S}_{\rm ct}
\ead 
where ${\cal S}_{\rm ct}$ corresponds to the counterterms introduced in the previous section. Then, the wavefunctional \eqref{Wavefunctional_e1} becomes
\bad \label{WilsonPsi}
\Psi [\phi_0,0] = \int {\cal D} \phi_c  \ \Psi_{\rm IR} [\phi_c,\t_c] \ \Psi_{\rm UV} [\phi_c, \t_c ; \phi_0,0]
\ead
with $\phi_0$ the limit of $\phi(\epsilon,\tx)$ for $\epsilon = 0$, $\phi_c = \phi(\t_c)$ and $\Psi_{\rm IR,UV}$ the parts of the wavefunctionals determined by $|\t| > |\t_c|$ and $|\t| < |\t_c|$ correspondingly. The main postulate of the Wilsonian procedure is that after integrating out the UV degrees of freedom, $\Psi_{\rm IR}$ will correspond to the generating functional of a dual QFT desribed by an action ${\cal S}_0$ of boundary fields $M$ and possessing a UV cutoff $\Lambda$:
\bad \label{IRPsi}
\Psi_{\rm IR} [\phi_c,\t_c] = \int {\cal D} M_{\tk < \Lambda} \exp \left\{ {\cal S}_0 +  \int d^3 \tx \ J(x) {\cal O} (x) \right\} \ . 
\ead  
Following the discussion in the previous subsection, we read the source from the boundary value of the bulk field so that in momentum space $J_\tk = H^{-1}_0 \phi_{c;\tk}  |\t_c|^{-\Delta_\mp}$ with the $\mp$ sign depending on the choice between the standard or alternative quantization. From now on, the upper, lower signs will denote the former and latter respectively. On the other side, the UV wavefunctional is 
\bad
\Psi_{\rm UV} [\phi_c, \t_c ; \phi_0,0] \simeq \exp \left\{-  \left[ \int^0_{\t_c} d \t \ \sqrt{-g} {\cal L} + {\cal S}_{\rm ct} + {\cal S}_{\rm bd}  \right] \right\}
\ead
with ${\cal S}_{\rm bd}$ a Gibbons-Hawking like term for the matter field that ensures a well-defined variational principle. The usual procedure is to make the quadratic ansatz:
\bad \label{Sbd}
{\cal S}_{\rm bd} = {\cal C} (\t_c) + \frac{1}{2} \int d^3 \tk \sqrt{-\gamma} \left\{ G (\t_c, \tk) \phi_{c,-\tk}  + \phi_{c,\tk} F (\t_c, \tk) \phi_{c,-\tk} \right\}
\ead 
and use the Hamilton-Jacobi formalism which is reviewed in \sect{HJformalism} in order to reach the flow equations \eqref{flow eq.} for $F(\t_c ,\tk), G(\t_c ,\tk)$ and $C(\t_c)$. Once we perform the $\phi_c$ path integral \eqref{WilsonPsi}, we will obtain the Wilsonian effective action ${\cal S}_{\rm eff}$ which will contain a single and double trace deformation that move the boundary theory away from the original CFT \eqref{CFTpart}: 
\bad \label{Seff}
\exp\left\{- {\cal S}_{\rm eff}\right\}  \propto  \exp \left\{ - \int d^3 \tk \ \left[ g_{-\tk}(\t_c) {\cal O}_\tk + f (\t_c, \tk) {\cal O}_\tk {\cal O}_{-\tk} \right] \right\} \ . 
\ead
Then, the coefficients of \eq{Sbd} will be directly associated with the dimensionless couplings $\tilde g_{-\tk}(\t_c) = |\t_c|^{- \Delta_\pm} g_{-\tk}(\t_c)$, $\tilde f (\t_c, \tk) = |\t_c|^{\mp 2 \n} f (\t_c, \tk)$:
\bad \label{fF}
\tilde g_{-\tk } (\t_c) =  |\t_c|^{-3} H^{-1}_0 \frac{G(\t_c, \tk)}{F(\t_c, \tk)}, \qquad \tilde f (\t_c, \tk) = |\t_c|^{-3} \frac{H^{-2}_0}{\sqrt{-\gamma} F(\t_c, \tk)}
\ead
and from \eq{flow eq.} we can retrieve their beta-functions up to first order in $H^2_0/M^2_{\rm pl}$
\begin{subequations} 
\bad \label{betag}
\beta_{\tilde g} \equiv  -|\t_c| \partial_{|\t_c|} \tilde g_{-\tk } (\t_c)  =\left\{ 3-  \left[ \left(1 + 2 {\cal C} (m,\xi) \frac{H^3_0}{M^2_{\rm pl}}  |\t_c| \right) |\t_c|^2 \tmk^2 + \left(1 + 4 {\cal C} (m,\xi) \frac{H^3_0}{M^2_{\rm pl}}  |\t_c| \right)\frac{\m^2}{H^2_0} \right]    \tilde f  (\t_c, \tk) \right\}  \tilde g_{-\tk } (\t_c)
\ead
and
\bad \label{betaf}
\beta_{\tilde f} \equiv  -|\t_c| \partial_{|\t_c|} \tilde f (\t_c, \tk) 
&= - \left[ \left(1 + 2 {\cal C} (m,\xi) \frac{H^3_0}{M^2_{\rm pl}}  |\t_c| \right) |\t_c|^2 \tmk^2 + \left(1 + 4 {\cal C} (m,\xi) \frac{H^3_0}{M^2_{\rm pl}}  |\t_c| \right)\frac{\m^2}{H^2_0} \right] \tilde  f^2 (\t_c, \tk) \\
& \qquad + 3  \tilde f (\t_c, \tk)  -  \left(1 - 2 {\cal C} (m,\xi) \frac{H^3_0}{M^2_{\rm pl}}  |\t_c| \right)
\ead
\end{subequations}
with the latter obtaining a Ricatti form. As a consistency check, we can momentarily discard $|\t_c|^2 \tmk^2$ so that we effectively decouple the non-zero momentum case and solve for the fixed point $\beta_f =0$. At $\t_c = 0$, this leads to the fixed point:
\bad \label{f*}
\tilde f^* = \frac{1}{\Delta_\mp} \ . 
\ead 
which we can substitute back into \eq{fF} and find $F(0,\tk) = \left(\frac{3}{2} \mp \n \right) H_0$. Thus, we are able to effectively retrieve the first counterterm \eqref{c_1} as long as we choose the upper sign. Of course, this is expected since placing the cutoff at $\t_c = 0$ means that no UV modes are integrated out and the original theory should be recovered. Note that in the pure dS case $({\cal C} (m,\xi) =0)$, the upper sign corresponds to  $|\t_c|= 0$ while the lower to the $|\t_c| \rightarrow \infty$ limit.  

Although both $\tilde f, \tilde g$ are independent of whether we impose the standard or alternative quantization as a boundary condition at the future boundary, our choice fixes $g \sim |\t_c|^{-\Delta_{\mp}}$ up to leading order in $|\t_c|$. Consequently, the single trace operator ${\cal O}$ will be of classical scaling dimension $3 - \Delta_\mp  = \Delta_\pm$ as expected and ${\cal O}^2$ will be irrelevant (upper sign) or relevant (lower sign) since $2 \Delta_+ > 3$ and vice versa. However, if we also consider the quantum corrections, the engineering dimension of ${\cal O}$ is $[{\cal O}] = \Delta^{\rm cl}_{{\cal O}} + \gamma_{\cal O}$, where $ \Delta^{\rm cl}_{{\cal O}} = \Delta_\pm$, $\gamma_{\cal O} \equiv  \mu \frac{d}{d \m} \ln {\cal O}$ denote the operator's classical and anomalous dimension respectively. Following the usual Wilson prescription, we require that \eq{Seff} retain its form under coarse-graining. This means that the total anomalous dimensions of the single and double trace operators near a fixed point $g^*,f^*$ can be approximated by $\gamma_{\cal O} = \frac{d \beta_g}{d g } \bigl|_{\tilde g = \tilde g^*}$ and $\gamma_{{\cal O}^2} = \frac{d \beta_f}{d f} \bigl|_{\tilde f = \tilde f^*}$ respectively. 

Clearly, in order to extract $\gamma_{\cal O}$ and $\gamma_{{\cal O}^2}$ from \eq{betag} and \eqref{betaf}, we ought to express the latter as a powerseries with respect to each dimensionless coupling. Near $\t_c =0$, we can write the coupling as a sum of its fixed point value \eqref{f*} plus a variation $\delta \tilde f$:
\bad \label{tildefexp}
\tilde f=  \tilde f^* + \delta \tilde f, \qquad \delta \tilde f = - \sum^{\infty}_j f_j H^j_0 |\tau_c|^j 
\ead
where the latter can be expanded as a power series wrt to $|\t_c|$. Here, it is important to point out that in the pure dS case, the leading order solution is ${\cal O}(|\t_c|^{2 \n})$. However, the emergent ${\cal O} (|\t_c|)$ terms in the beta-function dominate and end up forcing the expansion \eqref{tildefexp}. Then, we plug the above into the beta-function \eqref{betaf} and up to second order in $|\t_c|$, we find by power-counting:
\begin{subequations}
\bad
f_1 = \frac{2 {\cal C} (m,\xi) H^2_0 }{(1 - 2 \nu)M^2_{\rm pl}} \left(1 - 2\frac{\Delta_+}{\Delta_-} \right)
\ead
and
\bad
f_2 = \frac{- \Delta_+ \Delta_- f^2_1 + 8 \Delta_+ f_1 {\cal C} (m,\xi) \frac{H^2_0}{M^2_{\rm pl}} }{2(1 -  \nu)} \ . 
\ead  
\end{subequations}
Notice the $\nu=\frac{1}{2}$ and $\nu=1$ poles for each coefficient. However, the former is simplified when ${\cal C} = {\cal C}_\beta$, since ${\cal C}_\beta \rightarrow \frac{\pi}{2} (1 - 2 \n) $ in the $\nu \rightarrow 1/2$ limit, while the latter corresponds to strong bacreactions effects \cite{FotisAntonis3} that exceed the scope of this paper. 

Moving on, we reverse engineer \eq{tildefexp} in terms of $|\tau_c|$ up to second order and substitute the result in \eq{betaf}. We find
\bad \label{betaf2}
\beta_{\tilde f} = - \delta \tilde f  - \frac{\left(\frac{3}{2} + \n \right) \left( \frac{3}{2} - \n \right)}{6(1 - \nu)} \frac{11 - 10 \n}{1+ 2 \n }  \delta \tilde f^2  
\ead 
and the engineering dimension of ${\cal O}^2$:
\bad
[ {\cal O}^2 ] = 2 \Delta_\pm + \frac{d \beta_f}{d f}  = 3 + \frac{d \beta_{\tilde f}}{d \tilde f}  = 2 - \frac{\left(\frac{3}{2} + \n \right) \left( \frac{3}{2} - \n \right)}{3(1 - \nu)} \frac{11 - 10 \n}{1+ 2 \n }   \delta \tilde f \ .
\ead  
Interestingly, this type of behavior is reminiscent of the $O(N)$ vector $\sigma$ model \footnote{here $\sigma^i$ is the scalar vector field of the dual theory.} in $4 - \delta$ dimensions. In particular, the dimensional regularization (as reviewed in Sect. 4 of \cite{Henriksson}) produces for $\delta = 1$ a similar beta-function as \eq{betaf2} and near the UV fixed point the engineering dimension $[\sigma^4] = 2 + {\cal O} ({\cal \delta} g)$. Actually, a connection between the dS/CFT duality and the $Sp(N)$ vector models has already been proposed by \cite{ Anninos2013, Das2013}. The latter can be thought of as an analytic continuation of an $O(N)$ model in the Large N limit corresponding to a negative $N$. This also explains the opposite sign of the second order contribution in \eq{betaf2} in the $\n<1$ and $\n > 1.1$ regimes. In particular, \cite{Das2013} assumed a deformed CFT in the alternative quantization via a double-trace operator of dimension $2\Delta_-$ and computed its corresponding beta-function. However, for arbitrary $\n$ these kind of composite operators cannot be built from a local QFT. In contrast, our analysis showcases that the introduction of linear contributions into the flow equations can alter the resulting RG equations so that in leading order, a duality between the effective action \eqref{Seff} and a ${\cal O} (N)$ or $Sp(N)$ is further strengthened.    

\section{Conclusions}

Throughout this work, we consider a spacetime that is slightly deformed due to a semiclassical thermal backreaction of the initial Poincare patch of the de Sitter background. Accordingly, it constitutes a direct continuation of our previous study \cite{FotisAntonis3}. There, a thermal state characterized by the de Sitter temperature was considered, whose origin could be either an explicit thermal bath or the coarse-graining associated with horizon crossing of modes. The resulting spacetime retains the FLRW form, with a scale factor receiving a constant first-order correction that, in turn, brings the original bulk equation of motion into Whittaker form.

The goal of this paper was to explore the physical implications of the corresponding modes in two key applications of de Sitter spacetime, namely inflation and holography. Using canonical quantization, each mode is expressed as a linear combination of the two Whittaker functions, from which we derive the normalization condition that constrains their coefficients. The latter can be understood as a generalization of its de Sitter analogue, fixed by a nonzero parameter, and is required to match the Bunch–Davies condition at future infinity.

At leading order, the deviation produces a contribution to the de Sitter Hubble parameter that is linear in conformal time. This, in turn, yields a first flow parameter $\epsilon_H$ at late times and a second flow parameter $\eta_H$ of order minus one. We interpreted this via a single field model and showed that its effective potential is quadratic and matches at first order the potential of a certain constant-roll model in the attractor regime. Accordingly, the co-moving curvature perturbations obtain an extra  contribution that freezes at the superhorizon limit and amplifies the power-spectrum at small scales. This type of UV enhancement is speculated to have occurred for a transient period of a few e-folds near the inflation end and resulted to the formation of primordial black holes. We briefly discussed that a distinction from the constant-roll model should happen at second order and will result to further fluctuations. 

At the future boundary, we utilized the dS/CFT postulate together with the new normalization condition and demonstrated that, in the complementary series, the coefficient of the CFT propagator takes a more general form. Following the Holographic normalization procedure, we argued that the thermal backreaction does not introduce additional logarithmic terms and computed the counterterms of the on-shell action at the boundary. The latter is treated as a UV fixed point and the introduced scale-dependence alters the flow-equation of the Wilsonian RG flow. Near this fixed point, we expressed the beta-function of the generalized double-trace coupling as a series in terms of its deviation from the fixed point value and calculated the corresponding anomalous dimension. Our results further support the proposed duality between a bulk quasi dS scalar model and a 3d boundary $Sp(N)$ theory.
 
\section*{Acknowledgments}
I thank my supervisor N. Irges, and L. Karageorgos for helpful discussions throughout this project. 
%%%%%%%%%%%%%%%%%%%%%%%%%%%%%%%%%%%%%%%%%%%%%%%%%%%%%%%%%%%%%

%%%%%%%%%%%%%%%%%%%%%%%%%%%%%%%%%%%%%%%%%%%%%%%%%%%%%%%%%%%%%%%%%%%%%%%%%%%%

\appendix

\section{Whittaker functions useful relations}
In this section of the appendix, we include all the useful properties and relations that the Whittaker functions satisfy, as given in \cite{DLMF}. The Whittaker differential equation for a function W(z):
\bad
\frac{d^2 W}{dz^2} + \left(-\frac{1}{4} + \frac{\kappa}{z} + \frac{\frac{1}{4} - \nu^2}{z^2} \right) W =0 
\ead
has the two standard solutions:
\begin{subequations}
\bad \label{Whittaker_sol_M}
M_{\kappa,\nu} (z) = e^{- \frac{1}{2} z} z^{ \nu + \frac{1}{2}} \ M \left(\frac{1}{2} +  \nu -  \kappa, 1 + 2  \nu, z\right) ,
\ead
\bad \label{Whittaker_sol_W}
W_{\kappa,  \nu} (z) = e^{- \frac{1}{2} z} z^{ \nu + \frac{1}{2}} \ U\left(\frac{1}{2} +  \nu - \kappa, 1 + 2  \nu, z\right) 
\ead
\end{subequations} 
which are themselves written in terms of the confluent hypergeometric fuctions. The former, contains Kummer's function:
\bad \label{Kummer_sol}
M(a,b,z) = \sum^\infty_{ n =0 } \frac{(a)_n}{(b)_n} \frac{z^n}{ n!} \ ,
\ead
while the latter is defined via the Tricomi function 
\bad \label{Tricomi_sol}
U(a,b,z) = \frac{1}{\Gamma (a)} \int^\infty_0 dt \ e^{-z t} t^{a-1} (1+t)^{b-a-1} \ .
\ead 
The $M_{\kappa,\nu} (z)$ satisfies the analytic-continuation formula:
\bad \label{Analytic_cont M}
M_{\kappa,  \n} (z e^{\pm \pi i})  = \pm i e^{\pm  \n \pi i } M_{-\kappa,  \n} (z)  \ . 
\ead
Some of the known Wronskians between the Whittaker functions are
\begin{subequations} \label{Whittaker_Wronsk}
\bad 
W \left[M_{\kappa, \nu} (z)\ , W_{-\kappa,  \nu} (z e^{\pm \pi i})\right] =   \frac{\Gamma (1 + 2 \nu)}{\Gamma \left(\frac{1}{2} +  \n + \kappa \right)} e^{\mp \left(\frac{1}{2} +  \n \right) \pi i },
\ead
\bad
\tilde W \left[W_{\kappa,  \nu} (z)\ , W_{-\kappa,  \nu} ( e^{\pm \pi i } z)\right] =  e^{\mp \kappa \pi i} \ . 
\ead
\end{subequations}
Both Whittaker functions can be associated with the Bessel functions via the connecting formuilae:
\begin{subequations} \label{Whittaker_Bessel}
\bad
M_{0,\n} (z) = 2^{2 \n } \Gamma (1 + \n) \sqrt{z} I_\n \left( \frac{z}{2} \right),
\ead
\bad
W_{0,\n} (z) = \sqrt{\frac{z}{\pi}} K_\n \left( \frac{z}{2} \right)
\ead
\end{subequations}
In the small $|z| \rightarrow 0$ limit, the limiting forms of both functions are:
\begin{subequations} \label{Whittaker_Assympt}
\bad
M_{\kappa,\nu} (z) = z^{\nu + \frac{1}{2}}, \qquad 2 \nu \neq -1,-2,...
\ead
and
\bad
W_{\kappa,\nu} (z) = \frac{\Gamma (2 \nu)}{\Gamma \left( \frac{1}{2} + \nu - \kappa \right) } z^{\frac{1}{2} - \nu} +\frac{\Gamma (-2 \nu)}{\Gamma \left( \frac{1}{2} - \nu - \kappa \right) } z^{\frac{1}{2} + \nu} + {\cal O} \left( z^{\frac{3}{2} \pm \nu}\right)  \ . 
\ead
\end{subequations}

\section{Constant-roll inflation} \label{Constant-roll}
Constant-roll inflation is a family of single-filed models which satisfy the extra condition 
\bad
\frac{\ddot \varphi}{\dot \varphi} = \alpha H \ . 
\ead
Then, for a monotonic $t(\phi)$ we can write $\dot H = \dot \varphi \frac{d H}{d \varphi}$ and substitute it into the second Friedman \eq{2ndFried3}. Differentiating the result with respect to $t$ we obtain
\bad \label{CR Heq}
\frac{d^2 H}{d \varphi^2} = -\frac{1}{2 M^2_{\rm pl}} \alpha H 
\ead
which has the general solution 
\bad \label{Hvarphi}
H(\varphi) = {\cal C}_1 \cosh \left( \sqrt{\frac{ |\a|}{2}} \frac{\varphi}{M_{\rm pl}} \right)   + {\cal C}_2 \sinh \left( \sqrt{ \frac{|\a|}{2}} \frac{\varphi}{M_{\rm pl}} \right) \ . 
\ead 
In \cite{Starobinsky2018} each distinct solution is examined with the most interesting case given for ${\cal C}_2 = 0$. Otherwise, ${\cal C}_2 \neq$ can lead to $\ddot a <0$ i.e. a decelerating phase especially at late-times.  
Substituting \eq{2ndFried3} into \eq{1stFried3} along with \eq{Hvarphi}, we solve for the potential and find
\bad \label{CRVeff}
\frac{1}{2 M^2_{\rm pl}} V_{\rm eff} (\varphi) &= 3 H(\varphi)^2 - 2 M^2_{\rm pl} \left(  \frac{d H(\varphi)}{d \varphi} \right)^2  \\
&= \frac{1}{2} {\cal C}^2_1 \left[ 3 - \alpha + (3 + \alpha) \cosh^2\left( \sqrt{2 |\a|} \frac{\varphi}{M_{\rm pl}} \right) \right] \ . 
\ead 
Notice that for $\alpha = -3$, the potential becomes constant which is the case of ultra-slow roll. 

The Mukhanov-Sasaki equation arising from the co-moving curvature perturbation $\cal R$ is \cite{Mukhanov1985, Sasaki1986}:
\bad \label{Mukhanov-Sassaki2}
v_\tmk ''  + \left(\tmk^2  - \frac{A''}{A} \right) v_\tmk = 0 \ . 
\ead
with $v_\tmk = A {\cal R}_\tmk , \  A^2 = 2 M^2_{\rm pl} a^2 \epsilon_H$ and 
\bad
\frac{A''}{A} \sim \frac{\n^2 - \frac{1}{4}}{\t^2}, \quad \nu = |\frac{3}{2} + \alpha| \ .
\ead
In the superhorizon limit $\tmk^2 << A''/A$, we can express \eq{Mukhanov-Sassaki2} wrt to {\cal R} which has the general solution
\bad \label{solutionR}
{\cal R}_\tmk = A_\tmk + \int dt \frac{B_\tmk}{a^3 \epsilon_H} 
\ead 
with $A_\tmk, B_\tmk$ arbitrary constants in time. At leading order, 
\bad
\epsilon_H \equiv - \frac{\dot H}{H^2} = \frac{1}{2 M^2_{\rm pl}} \left( \frac{\dot \varphi}{H}\right)^2 \propto a^{ 2\alpha}
\ead
so that the integrand of the second term in \eq{solutionR} scales with $a^{- (3+2 \a)}$. Evidently, an important distinction occurs at $\alpha = - \frac{3}{2}$. In particular, for $\alpha < - \frac{3}{2}$ the second term grows at late times resulting to a non-attractor solution. In contrast, for $\alpha > - \frac{3}{2}$ the curvature modes remain frozen in the superhorizon limit. 

Finally, using the definition of the scalar power-spectrum \eqref{PR2}, we calculate the scalar spectral index
\bad
n_S - 1 \equiv \frac{d \ln {\cal P_R}}{d \ln \tmk} = 3 - 2\n = 3 - |3 + 2 \alpha| 
\ead
which describes a blue-tilted spectrum in the $-\frac{3}{2} < \alpha < 0$ regime and red-tilted otherwise.

\section{The Hamilton-Jacobi equation} \label{HJformalism}
Consider the UV wave functional:
\bad \label{UVPsi}
\Psi_{\rm UV} [\phi_c, \t_c ; \phi_0,0] \simeq \exp \left\{-  \left[ \int^0_{\t_c} d \t \ \sqrt{-g} {\cal L} + {\cal S}_{\rm ct} + {\cal S}_{\rm bd}  \right] \right\}
\ead
with ${\cal L}$ representing the on-shell Lagrangian obtained from the classical solution and ${\cal S}_{\rm bd} [\phi_c, \t_c]$ a Gibbons-Hawking like term for the matter field that ensures a well-defined variational principle and encodes the data at the cut-off. Specifically, in order for the requirement $\delta {\cal S} / \delta \phi = 0$ to give the classical eom, we find the boundary condition:
\bad \label{bc1}
\pi_c = \frac{\delta {\cal S}_{\rm bd}}{\delta \phi_c}  
\ead 
with $\pi_c$ the canonical momentum of the bulk field at $\t = \t_c$. 

Consequently, the Wilsonian effective action is found by performing the path integral \eqref{WilsonPsi} with respect to $\phi_c$:
\bad \label{eff_action}
\exp\left\{-  {\cal S}_{\rm eff}\right\}  = \int {\cal D} \phi_c \exp \left\{ -\left[  {\cal S}_{\rm bd} [\phi_c, \t_c] - \int d^3 x J(x) {\cal O}  \right]\right\} 
\ead
and since the on-shell part of \eq{UVPsi} is fixed by the classical solution, we can disregard its contribution to the above path integral. 
 Of course, the dual theory should be independent from the choice of $\t_c$:
\bad
\frac{d}{d\t_c} Z = \frac{d}{d\t_c} \Psi [\phi_\epsilon, \epsilon] = \frac{d}{d\t_c} \braket{e^{- S_{\rm eff}}} = 0 
\ead
where the brakets denote the path integral \eqref{IRPsi}. This requirement is satisfied by the Schrodinger equations:
\bad
 \partial_{\t_c} \Psi_{\rm IR} [\t_c, \phi_{\t_c}] = -\hat {\cal H} (\hat \phi_{c},\hat \pi_{c}) \Psi_{\rm IR}[\t_c, \phi_{\t_c}]
\ead
\bad \label{ScrEqUV}
\partial_{\t_c} \Psi_{\rm UV} [\phi_c, \t_c ; \phi_0,0] =  \hat {\cal H} (\hat \phi_c,\hat \pi_c) \Psi_{\rm UV} [\phi_c, \t_c ; \phi_0,0]
\ead
which describe how each wavefunctional flows towards opposite directions. Here we promote the canonical momentum to the operator $\hat \pi_c = - \frac{\delta}{ \delta \phi_c}$ in the Schrodinger representation and 
\bad
\hat {\cal H} (\hat \phi_c,\hat \pi_c) = - \frac{1}{2}\int d^3 x  \left\{\frac{1}{a^2} \hat \pi^2_c + a^2 \delta^{ij} \partial_i \hat \phi_c \partial_j \hat \phi_c + a^4 \m^2 \hat \phi^2_c \right\}
\ead
the Hamiltonian operator that corresponds to \eq{Action} for a flat slicing choice of the original metric. Here, the Hamiltonian operator acts as the generator of time-evolution as usual and governs the flow of each wavefunctional as we push $\t_c$ away from the future boundary. In the semiclassical limit, the Fourier transform of \eq{ScrEqUV} leads to the Hamilton-Jacobi equation:
\bad \label{HJ}
 \partial_{\t_c} {\cal S}_{\rm bd} =   \frac{1}{2} \int d^3 \tk  \  \left\{\frac{1}{a^2} \frac{\delta {\cal S}_{\rm bd}}{\delta \phi_{c,\tk}} \frac{\delta {\cal S}_{\rm bd}}{\delta \phi_{c, -\tk}} +   a^2   \left(\tmk^2    + a^2\m^2 \right) \phi_{c,\tk} \phi_{c, -\tk}   \right\} 
\ead
where we have made use of the Fourier transform of the boundary condition \eqref{bc1}. Clearly, the above equation describes the flow of the Hamilton-Jacobi functional ${\cal S}_{\rm bd}$ as governed by the bulk mechanics and is a direct product of the boundary condition \eqref{bc1}.

The usual ansatz is to write ${\cal S}_{\rm bd}$ as a quadratic expansion of the Fourier transform $\phi_{c,\tk}$
\bad 
{\cal S}_{\rm bd} = {\cal C} (\t_c) + \frac{1}{2} \int d^d \tk \sqrt{-\gamma} \left\{ G (\t_c, \tk) \phi_{c,-\tk}  + \phi_{c,\tk} F (\t_c, \tk) \phi_{c,-\tk} \right\}
\ead
and plug it in \eq{HJ}. Grouping together the similar functional forms, we find the flow equations for the coefficients:
\begin{subequations} \label{flow eq.}
\bad
a^{-4} (\t_c)\partial_{\t_c} \left(\sqrt{-\gamma (\t_c)} F(\t_c, \tk) \right) =   F^2(\t_c , \tk) +  a^{-2} (\t_c) \tmk^2    + \m^2  
\ead
\bad
a^{-4} (\t_c) \partial_{\t_c} \left(\sqrt{-\gamma (\t_c)} G(\t_c, \tk) \right) = G(\t_c,\tk ) F(\t_c, \tk)
\ead
and
\bad
a^{-4} \partial_{\t_c} \left(\sqrt{-\gamma (\t_c) } {\cal C}(\t_c) \right) =  \int d^3 \tk \ G^2 (\t_c, \tk) \ . 
\ead
\end{subequations}
 
%%%%%%%%%%%%%%%%%%%%%%%%%%%%%%%%%%%%%%%%%%%%%%%%%%%%%%%%%%%%%%%%%%%%%%%%%%%%

%%%%%%%%%%%%%%%%%%%%%%%%%%%%%%%%%%%%%%%%%%%%%%%%%%%%%%%%%%%%%%%%%%%%%%%%%%%%

\printbibliography %prints bibliography

\end{document}